\documentclass[12pt,psfig]{article}
\usepackage{amssymb}
\usepackage{amsmath,amsthm}
\usepackage{amsfonts}
\usepackage{graphicx}
\usepackage{graphics}
\usepackage{indentfirst}
\numberwithin{equation}{section}

\usepackage{hyperref}

\begin{document}
\begin{center}\Large\textbf{Pair Production
of the Open Superstrings from the
Parallel-dressed D3-branes in the Compact Spacetime}
\end{center}
\vspace{0.75cm}
\begin{center}{\large Niloufar Barghi-Janyar and \large Davoud
Kamani}
\end{center}
\begin{center}
\textsl{\small{Department of Physics, Amirkabir University of
Technology (Tehran Polytechnic) \\
P.O.Box: 15875-4413, Tehran, Iran \\
e-mails: niloufarbarqi@aut.ac.ir , kamani@aut.ac.ir \\}}
\end{center}
\vspace{0.5cm}

\begin{abstract}

We employ the boundary state formalism to compute
the pair creation rate of the open superstrings
from the interaction of two parallel D3-branes.
The branes live in the partially compact spacetime.
In addition, they have been dressed with the
internal gauge potentials and the Kalb-Ramond field.

\end{abstract}


\newpage
\section{Introduction}

The D-branes are effective tools
to study the string theory \cite{1}.
For investigating the D-branes interaction
an appropriate approach is the boundary
state formalism \cite{2}-\cite{14}.
Since the boundary state encodes all
properties of the corresponding
D-brane, it is a powerful tool for studying
the various configurations of the D-branes.
For example, a phenomenon
similar to the Casimir effect, which is known as
the open string pairs creation from the branes system,
is one of these setups. The pair production of the open
strings in the bosonic and superstring
theories has been studied in various papers
\cite{15}-\cite{25}.
The fundamental strings creation through the
branes interaction demonstrates
the attractive force between the D-branes
\cite{19}-\cite{22}.

Using the boundary state formalism, we
shall compute the rate of the open superstring
pair creation from the interaction of two
parallel D3-branes. The branes have been
embedded in the partially compact spacetime.
Besides, the electric and magnetic
fields live on them. Therefore, at first we shall
compute the closed superstring amplitude
for a general branes dimension $p$, and then, we
rewrite the interaction amplitude for the case $p=3$.
For the small separation of the branes, the nature of the
interaction (repulsion or attraction)
completely is ambiguous. However, the
annulus open superstring amplitude is
an appropriate tool for describing such cases.
Thus, from the cylinder amplitude we shall obtain
the latter one. The one-loop annulus amplitude
enables us to extract the rate of the open
superstring pair creation. In fact, production 
of the superstring pairs
obviously indicates the decay of the underlying system.
As a special case, the pair production rate
for the particular electromagnetic
fields on the branes will be calculated.
We shall see that the compactification,
similar to the magnetic field,
drastically increases the pair production rate.

This paper is organized as follows. In Sec. 2,
the boundary state, associated with a partially compact
D$p$-brane in the presence of
the background fields, will be introduced.
In Sec. 3, the interaction
amplitude of a system, which consists of two
parallel-dressed D$p$-branes, will be calculated.
In Sec. 4, the decay rate of a system
of two D3-branes via the open superstring
pair production will be obtained.
Sec. 5 is devoted to the conclusions.

\section{The associated boundary state to the D$p$-brane}

We begin with the following sigma-model
action for the closed string
\begin{eqnarray}
S=&-&\frac{1}{4\pi\alpha'}\int_{\Sigma} {\rm d}^2\sigma
\left(\sqrt{-h} h^{ab}G_{\mu\nu} \partial_{a}X^\mu
\partial_b X^{\nu} +\varepsilon^{ab}
B_{\mu\nu}\partial_a X^{\mu}
\partial_b X^{\nu}\right)
\nonumber\\[10pt]
&+&\frac{1}{2\pi\alpha'}\int_{\partial\Sigma}
{\rm d}\sigma\left(A_{\alpha}
\partial_{\sigma}X^{\alpha}
\right).
\label{e:2.1}
\end{eqnarray}
We consider a constant Kalb-Ramond background
field $B_{\mu\nu}$. For the internal $U(1)$ gauge potential
$A_{\alpha}$, we apply the Landau gauge $A_{\alpha}=
-\frac{1}{2}F_{\alpha\beta}X^{\beta}$
whit the constant field strength $F_{\alpha\beta}$.
We employ the flat worldsheet $\Sigma$ with the metric
$h_{ab}=\eta_{ab}=\rm{diag}{(-1,1)}$,
which lives in the flat spacetime with the metric
$G_{\mu\nu}=\eta_{\mu\nu}=\rm{diag}{(-1,1,\cdots,1)}$.
Note that the devoted indices to the
D$p$-brane worldvolume directions are $\alpha,\beta
\in\{0,\alpha_{\rm n},\beta_{\rm n},
\alpha_{\rm c},\beta_{\rm c}\}$
with $\{\alpha_{\rm n},\beta_{\rm n}\}\cup
\{\alpha_{\rm c},\beta_{\rm c}\}=\{1,2,\cdots,p\}$.
Some of the brane directions and
some of the normal directions to it
are compacted on tori. Therefore,
the subscript ``n'' and ``c'' refer to the
non-compact and compact directions, respectively.
For the perpendicular directions
to the brane we shall use $i\in\{i_{\rm n},i_{\rm c}\}$
with $\{i_{\rm n}\}\cup \{i_{\rm c}\}=
\{p+1,\cdots,9\}$.

The bosonic portions of the boundary state equations
are obtained by the variation of above action with
respect to $ X ^{\mu} (\sigma, \tau)$,
\begin{eqnarray}
&~&\left(\partial_\tau X^\alpha
+\mathcal{F}^\alpha\;_{\beta}
\partial_{\sigma} X^{\beta}-B^\alpha\;_i
\partial_{\sigma} X^i
\right)_{\tau=0} |B_x\rangle=0,
\nonumber\\[7pt]
&~&\left(X^{i}-y^{i}\right)_{\tau=0}|B_x\rangle=0,
\label{e:2.2}
\end{eqnarray}
where $\mathcal{F}_{\alpha\beta}=
\partial_\alpha A_\beta-\partial_\beta
A_\alpha-B_{\alpha\beta}$ shows the
total field strength.
On the closed string boundary we applied
$\left(\delta X^i\right)_{\tau=0}
=\left(X^{i}-y^{i}\right)_{\tau=0}$.
The parameters $y^i$s represent the
location of the D$p$-branes. The second equation
implies that $\partial_\sigma X^i$ vanishes
on the boundary. Hence, the last term
of the first equation is removed.

The solution of the equation of motion of closed string is
\begin{eqnarray}
X^\mu(\sigma,\tau)=x^\mu+2\alpha'p^\mu\tau+2L^\mu\sigma+
\frac{i}{2}\sqrt{2\alpha'}\sum_{m\neq 0}\frac{1}{m}
\left(\alpha^\mu_m\;e^{-2im\left(\tau-\sigma\right)}+
\tilde{\alpha}^\mu_m\;e^{-2im
\left(\tau+\sigma\right)}\right).
\label{e:2.3}
\end{eqnarray}
For the non-compact and compact directions
$L^\mu$ is zero and $N^\mu R^\mu$, respectively.
If the $x^\mu$-direction is compact
it has the radius $R^\mu$. Thus, around this direction,
the closed string possesses a winding number $N^\mu$.
Note that we assume the time direction $x^0$
is non-compact.

By combining Eqs. \eqref{e:2.2} and
\eqref{e:2.3} the boundary state
equations take the features
\begin{eqnarray}
&~&\left[\alpha^\alpha_n +\tilde{\alpha}^\alpha_{-n}
-\mathcal{F}^\alpha\;_\beta\left(\alpha^\beta_n
-\tilde{\alpha}^\beta_{-n}\right)\right]|B_x\rangle=0,
\nonumber\\[7pt]
&~&\left(\alpha'p^{\alpha}+\mathcal{F}^{\alpha}\;_\beta
L^\beta\right)|B_x\rangle=0,
\nonumber\\[7pt]
&~&\left(\alpha^{i}_n
-\tilde{\alpha}^{i}_{-n}\right)|B_x\rangle=0,
\nonumber\\[7pt]
&~&\left(x^i-y^i\right)|B_x\rangle=0,
\nonumber\\[7pt]
&~&L^i|B_x\rangle=0.
\label{e:2.4}
\end{eqnarray}
The last equation implies that the closed string cannot
wind around the compact directions which are
perpendicular to the brane.

The action \eqref{e:2.1}
is invariant under the global worldsheet
supersymmetry. Therefore, by using the replacements
$\partial_+ X^\mu\left(\sigma,\tau\right)\to
i\eta\Psi^\mu_+\left(\sigma,\tau\right)$ and
$\partial_- X^\mu\left(\sigma,\tau\right)\to
-\Psi^\mu_-\left(\sigma,\tau\right)$
in the bosonic Eqs. \eqref{e:2.2} we acquire the
fermionic parts of the boundary state equations
\begin{eqnarray}
&~&\left[\left(\Psi^\alpha_{-}
-\mathcal{F}^\alpha\;_{\beta}\Psi^\beta_{-}\right)
-i\eta\left(\Psi^\alpha_{+}
+\mathcal{F}^\alpha\;_{\beta}\Psi^\beta_{+}
\right)\right]_{\tau=0} |B_\Psi,\eta\rangle=0,
\nonumber\\[10pt]
&~&\left(\Psi^i_{-}+i\eta\Psi^i_{+}\right)_{\tau=0}
|B_\Psi,\eta\rangle=0,
\label{e:2.5}
\end{eqnarray}
where $\eta=\pm1$ is preserved for applying the
GSO projection. In terms of the fermions oscillators
these boundary conditions take the following forms
\begin{eqnarray}
&~&\left(\Psi^\alpha_q -i\eta\mathcal{Q}^\alpha\;_\beta
\tilde{\Psi}^\beta_{-q}\right)|B_\Psi,\eta\rangle=0,
\nonumber\\[10pt]
&~&\left(\Psi^i_q+i\eta\tilde{\Psi}^i_{-q}\right)
|B_\Psi,\eta\rangle=0,
\label{e:2.6}
\end{eqnarray}
where the index $q$ is integer (half-integer) for the R-R
(NS-NS) sector. The orthogonal matrix
$\mathcal{Q}$ possesses the definition
$\mathcal{Q}=\left({\bf 1}-\mathcal{F}\right)^{-1}
\left({\bf 1}+\mathcal{F}\right)$.

Now we solve Eqs. (\ref{e:2.4}) and (\ref{e:2.6}).
For the bosonic part, the solution of
equations (\ref{e:2.4}) is given by
\begin{eqnarray}
|B_x\rangle &=&
\frac{T_p}{2}\sqrt{\det\left({\bf 1}-\mathcal{F}\right)}
\prod_i\delta\left(x^i-y^i\right)|p^i=0\rangle
\nonumber\\
&\times& \exp{\left(-\sum^\infty_{m=1}
\frac{1}{m}\alpha^\mu_{-m}
\mathcal{S}_{\mu\nu}\tilde{\alpha}^\nu_{-m}\right)}
|0\rangle_\alpha |0\rangle_{\tilde \alpha}
\prod_\alpha|p^\alpha\rangle,
\label{e:2.7}
\end{eqnarray}
in which $\mathcal{S}^\mu\;_\nu=
\left(\mathcal{Q}^\alpha\;_\beta \;, -\delta^i\;_j\right)$,
and $T_p$ refers to the tension of the D$p$-brane.
The factor $\sqrt{\det\left({\bf 1}-\mathcal{F}\right)}$
is originated from the path integral in
the presence of the Gaussian boundary action
\cite{26}-\cite{28}.

Eqs. \eqref{e:2.6} give the boundary state of
the NS-NS and R-R sectors as in the following
\begin{eqnarray}
|B_\Psi, \eta\rangle_{\rm NS} &=&
\exp\left(i\eta\sum_{r=1/2}^{\infty}
b^\mu_{-r}\mathcal{S}_{\mu\nu} \tilde{b}^\nu _{-r}
\right)|0\rangle,
\label{e:2.8}
\end{eqnarray}
\begin{eqnarray}
|B_\Psi, \eta\rangle_{\rm R} &=&
\left[\det\left({\bf 1}-\mathcal{F}\right)\right]^{-1/2}\;
\exp\left(i\eta\sum_{n=1}^{\infty} d^\mu_{-n}
\mathcal{S}_{\mu\nu} \tilde{d}^\nu _{-n} \right)
|B_\Psi,\eta\rangle^{(0)}_{\rm R}.
\label{e:2.9}
\end{eqnarray}
In comparison with the bosonic equation \eqref{e:2.7},
the factor $\sqrt{\det\left({\bf 1}-\mathcal{F}\right)}$
has been reversed. This is because of the presence of the
Grassmannian variables in computing the path integral.
The zero-mode part of the R-R sector possesses the form
\begin{eqnarray}
|B_\Psi, \eta\rangle^{(0)}_{\rm R}=
\left(C\Gamma^0\Gamma^1\cdots\Gamma^p \;
\frac{1+i\eta\Gamma_{11}}{1+i\eta}U\right)_{AB}
|A\rangle|\tilde{B}\rangle,
\end{eqnarray}
where A and B represent
the 32-dimensional indices for the spinors,
$\Gamma^\mu s$ show the Dirac matrices in the
10-dimensional spacetime, $|A\rangle|\tilde{B}\rangle$
is the vacuum of the R-R zero modes $d^\mu_0$ and
$\tilde{d}^\mu_0$, $C$ defines the charge
conjugation matrix, and the matrix $U$ has the
definition
\begin{eqnarray}
U=;\exp\left(\frac{1}{2}\mathcal{F}_{\alpha\beta}
\Gamma^\alpha\Gamma^\beta\right); .
\end{eqnarray}
The conventional notation $;\;;$ implies that
we should expand the exponential with the
convention that all $\Gamma$-matrices
anticommute. Thus, for each value of $p$,
we receive a finite number of terms.

The total boundary state, corresponding to the D$p$-brane,
in each sector is given by the product of the
matter- and ghost-parts
\begin{equation}
|B,\eta\rangle_{\rm NS(R)}=|B_x\rangle
|B_\Psi,\eta\rangle_{\rm NS(R)}
|B_{\rm{gh}}\rangle|B_{\rm{sgh}},
\eta\rangle_{\rm NS(R)}.
\label{e:2.12}
\end{equation}
The GSO-projected boundary states in the NS-NS
and R-R sectors have the forms
\begin{equation}
|B\rangle_{\rm NS}=\frac{1}{2}\left[|B,+\rangle_{\rm NS}
-|B,-\rangle_{\rm NS}\right],
\quad\;\;\;
|B\rangle_{\rm R}=\frac{1}{2}\left[|B,+\rangle_{\rm R}
+|B,-\rangle_{\rm R}\right].
\end{equation}
Since the conformal ghosts and
super-conformal ghosts are
not influenced by the background fields
we shall apply the standard
form of their boundary states.

\section{The D$p$-branes interaction}

The overlap of the boundary states,
via the closed string propagator,
gives the tree-level interaction amplitude
between two parallel-dressed branes 
$\mathcal{A}_{\rm NS-NS(R-R)}
= _{\rm NS(R)}\langle B_1|D|B_2\rangle_{\rm NS(R)}$.
The closed string propagator is as follows
\begin{eqnarray}
D & = &\frac{\alpha'}{4\pi}\int_{|z|\le 1}
\frac{{\rm d}^2z}{|z|^2}z^{L_0}\bar{z}^{\bar{L}_0},
\end{eqnarray}
where $L_0$ and $\tilde{L}_0$ are
the right- and left-moving total Virasoro
generators, including the
bosonic, fermionic, conformal ghosts and
super-conformal ghosts parts.
The total interaction amplitude, via the exchange of
the closed superstring, is the cylinder amplitude
$\mathcal{A}_{\rm{cylinder}}
=\mathcal{A}_{\rm NS-NS}+\mathcal{A}_{\rm R-R}$.
Thus, we acquire
\begin{eqnarray}
\mathcal{A}_{\rm{cylinder}}
&=& \frac{2^{\left[\frac{p+1}{2}\right]-3}\;T_p^2}
{\left(2\pi\right)^{9-p}}\alpha'
\left[\det{\left({\bf 1}-\mathcal{F}_1\right)}
\det{\left({\bf 1}-\mathcal{F}_2\right)}\right]^{1/2}
\int^\infty_0{\rm d}t\bigg\{\left(\frac{\pi}
{\alpha' t}\right)^{d_{i_{\rm n}}/2}
\nonumber\\
&\times&\exp{\left(-\frac{ Y^2_{\rm n}}{4\alpha' t}\right)}
\prod_{i_{\rm c}}\Theta_3\left(\frac{y_{1}^{i_{\rm c}}
-y_{2}^{i_{\rm c}}}
{2\pi R_{i_{\rm c}}}\bigg{|}\frac{i\alpha' t}
{\pi R_{i_{\rm c}}^2}\right)
\left[\eta\left(it\right)
\right]^{3\left[\frac{p+1}{2}\right]-12}
\nonumber\\
&\times& \left(\sum_{\{N^{\alpha_{\rm c}}\}}
\bigg[\left(2\pi\right)^{p+1}
\prod_{\alpha}\delta\left(p^\alpha_1-p^\alpha_2\right)
\exp{\left(-\frac{t}{\alpha'}l^{\alpha_{\rm c}}
l^{\beta_{\rm c}}\left(\delta_{\alpha_{\rm c}\beta_{\rm c}}
+\mathcal{F}^{\alpha}_{1\;\alpha_{\rm c}}
\mathcal{F}_{2\;\alpha\beta_{\rm c}}\right)\right)}\bigg]
\right)
\nonumber\\[7pt]
&\times&\bigg(-\Theta_2(0|it)^{4-[(p+1)/2]}
\prod^{[\frac{p-1}{2}]}_{a=0}\Theta_2(\nu_a|it)
+\Theta_3(0|it)^{4-[(p+1)/2]}\prod^{[\frac{p-1}{2}]}_{a=0}
\Theta_3(\nu_a|it)
\nonumber\\[7pt]
&\times&-\Theta_4(0|it)^{4-[(p+1)/2]}
\prod^{[\frac{p-1}{2}]}_{a=0}
\Theta_4(\nu_a|it)\bigg)
\prod^{[\frac{p-1}{2}]}_{a=0}
\frac{\sin(\pi \nu_a)}{\Theta_1(\nu_a|it)}
\bigg\},
\label{e:3.2}
\end{eqnarray}
where 
$l^{\alpha_{\rm c}}=N^{\alpha_{\rm c}} R^{\alpha_{\rm c}}$
is the eigenvalue of the operator $L^{\alpha_{\rm c}}$, and
$d_{i_{\rm n}}$ is the dimension of the transverse
non-compact directions.
The transverse compact directions induce the
$\Theta_3$-function in the second line. Besides,
the $\Theta_2$-function comes from the R-R sector,
and the $\Theta_3$- and $\Theta_4$-functions
originate from the NS-NS sector.
In addition, the quantities $\{e^{2\pi i \nu_a}|a = 1, 2,
\cdots, [(p-1)/2]\}$ are the
eigenvalues of the orthogonal matrix
$\Lambda=Q_1^{\rm T}Q_2$. They obviously depend
on the magnetic and electric fields on the branes.
The integer part of $(p \pm 1)/2$ 
is denoted by $[(p \pm 1)/2]$.
The branes distance along the non-compact
directions is $Y_{\rm n}=\left[\sum_{i_{\rm n}}
\left(y^{i_{\rm n}}_1-y^{i_{\rm n}}_2\right)^2
\right]^{1/2}$. Note that in the third 
line, the momentum components are extracted from the second
equation of Eq. \eqref{e:2.4}, i.e.,
$p^\alpha_{1(2)}=-\frac{1}{\alpha'}
\mathcal{F}^\alpha_{1(2)\;\beta_{\rm c}}
N^{\beta_{\rm c}}R^{\beta_{\rm c}}$.

From now on, for simplification, we restrict our setup
to the case $p=3$. This dimension also is
appropriate for relating the D-branes to the
real world. However, the amplitude for two parallel
D3-branes takes the feature
\begin{eqnarray}
\mathcal{A}_{\rm{cylinder}}
&=&\frac{ T_3^2 \alpha'}{(2\pi)^{6}}
\sqrt{\det ({\bf 1}-\mathcal{F}_1)\det({\bf 1}-\mathcal{F}_2)}
\int^\infty_0{\rm d} t\bigg\{\left(\frac{\pi}{\alpha' t}
\right)^{d_{i_{\rm n}}/2}
\nonumber\\
&\times&\exp{\left(-\frac{Y^2_{\rm n}}
{4\alpha' t}\right)}\prod_{i_{\rm c}}
\Theta_3\left(\frac{y_{1}^{i_{\rm c}}
-y_{2}^{i_{\rm c}}}{2\pi R_{i_{\rm c}}}\bigg{|}
\frac{i\alpha' t}{\pi R^2_{i_{\rm c}}}\right)
\left[\eta (i t )\right]^{-6}
\nonumber\\
&\times&\left(\sum_{\{N^{\alpha_{\rm c}}\}}
\left(2\pi\right)^4\prod_\alpha
\delta(p^\alpha_1-p^\alpha_2)
\exp{\left(-\frac{t}{\alpha'}l^{\alpha_{\rm c}}
l^{\beta_{\rm c}}\left(\delta_{\alpha_{\rm c}
\beta_{\rm c}}+
\mathcal{F}^{\alpha}_{1\;\alpha_{\rm c}}
\mathcal{F}_{2\;\alpha\beta_{\rm c}}\right)
\right)}\right)
\nonumber\\
&\times&\left[\;\Theta^2_1\left(\frac{\nu_0+\nu_1}{2}
\bigg{|}i t\right)
\;\Theta^2_1\left(\frac{\nu_0-\nu_1}{2}
\bigg{|}i t\right)\right]
\prod^{1}_{a=0}\frac{\sin{\left(\pi\nu_a
\right)}}{\Theta_{1}\left(\nu_a |i t\right)}\bigg\},
\label{e:3.3}
\end{eqnarray}
in which we employed the following identity for the
theta-functions \cite{29},
\begin{eqnarray}
&-\Theta_{2}\left(0|i t\right)^2
\Theta_{2}\left(\nu_0|i t\right)
\Theta_{2}\left(\nu_1|i t\right)
+\Theta_{3}\left(0|i t\right)^2
\Theta_{3}\left(\nu_0|i t\right)
\Theta_{3}\left(\nu_1|i t\right)
\nonumber\\[10pt]
&-\Theta_{4}\left(0|i t\right)^2
\Theta_{4}\left(\nu_0|i t\right)
\Theta_{4}\left(\nu_1|i t\right)
=2\Theta^2_1(\frac{\nu_0+\nu_1}{2}|i t)
\Theta^2_1(\frac{\nu_0-\nu_1}{2}|i t).
\label{e:3.4}
\end{eqnarray}

According to the orthogonality
of the matrix $\Lambda$ and the characteristics
of its eigenvalues, one of the $\nu_a$s
is pure imaginary, e.g. $\nu_0=i\upsilon_0$.
The range of the real $\nu_a$s and
$\upsilon_0$ are $\nu_a\in[0,1)$ and
$\upsilon_0\in(0,\infty)$, respectively.
Now consider small distance of the branes
(which is corresponding to the small value of ``$t$'').
Thus, the infinite product form of
$\Theta_1 (\nu_a|it)$ in the denominator of Eq. \eqref{e:3.3}
gives the factor 
$\prod^\infty_{n=1}[1-\cosh(2\pi \upsilon_0)]$.
Therefore, the sign of the amplitude will be
proportional to $(-1)^\infty$, which is ambiguous.
This demonstrates that the nature of the interaction
(repulsion or attraction) obviously is ambiguous.
This ambiguity indicates a new phenomenon.
Precisely, the exponential factor
$\exp{\left(-Y^2_{\rm n}/4\alpha' t\right)}$
for the small $t$ and also the foregoing ambiguity
play a significant role for occurrence
of this new phenomenon.
In fact, this new phenomenon implies the decay of the
system via the open superstring pair production.

\section{The rate of open superstring pair creation}

The Jacobi transformation $t\to 1/t'$ will be
used to calculate the open superstring one-loop interaction
amplitude from the cylinder one. This
enables us to understand the nature of the foregoing
physical phenomenon. In fact, this new phenomenon
comprises the open superstring
pair production, which we will be found later.
However, the annulus amplitude possesses the feature
\begin{eqnarray}
\mathcal{A}_{\rm annulus}
&=&\frac{T_3^2 \pi^{d_{i_{\rm n}}/2}}{(2\pi)^{6}
\alpha'^{(d_{i_{\rm n}}-2)/2}}
\sqrt{\det ({\bf 1}-\mathcal{F}_1)\det({\bf 1}-\mathcal{F}_2)}
\int^\infty_0{\rm d} t'\bigg\{
\nonumber\\
&\times&\exp{\left(-t'Y^2_{\rm n}/4\alpha'\right)}
t'^{(d_{i_{\rm n}}-8)/2}\;\prod_{i_{\rm c}}
\Theta_3\left(\frac{y_{1}^{i_{\rm c}}
-y_{2}^{i_{\rm c}}}{2\pi R_{i_{\rm c}}}\bigg{|}
\frac{i\alpha'}{t'\pi R^2_{i_{\rm c}}}\right)
\nonumber\\
&\times&\left[ \sum_{\{N^{\alpha_{\rm c}}\}}
\left(2\pi\right)^{4}\prod_\alpha
\delta(p^\alpha_1-p^\alpha_2)
\exp{\left(-\frac{1}{t'\alpha'}l^{\alpha_{\rm c}}
l^{\beta_{\rm c}}\left(\delta_{\alpha_{\rm c}\beta_{\rm c}}+
\mathcal{F}^{\alpha}_{1\;\alpha_{\rm c}}
\mathcal{F}_{2\;\alpha\beta_{\rm c}}\right)
\right)}\right]
\nonumber\\
&\times&\frac{\sin(\pi\nu_0)\sin(\pi\nu_1)}
{\sinh{(\pi\nu_0 t')}\sinh{(\pi\nu_1 t')}}
\left[\cosh{(\pi\nu_0 t')}-\cosh{(\pi\nu_1 t')}\right]^2
\prod^{\infty}_{m=1}\mathcal{C}_m (t') \bigg\},
\label{e:4.1}
\end{eqnarray}
where $\mathcal{C}_m (t')$ is given by
\begin{eqnarray}
\mathcal{C}_m (t') &=&
\frac{\left[1-2\cosh{(\pi t'(\nu_0+\nu_1))}
e^{-2 m\pi t'}+e^{-4 m\pi t'}\right]^2}
{{\left(1-e^{-2 m\pi t'}\right)^4}
\left[1-2\cosh{(2\pi t'\nu_0)}
e^{-2 m\pi t'}+e^{-4 m\pi t'}\right]}
\nonumber\\[7pt]
&\times&\frac{\left[1-2\cosh{(\pi t'(\nu_0-\nu_1))}
e^{-2 m\pi t'}+e^{-4 m\pi t'}\right]^2}
{1-2\cosh{(2\pi t' \nu_1)}e^{-2 m\pi t'}
+e^{-4 m\pi t'}}.
\nonumber\\
\end{eqnarray}
To acquire this feature of
Eq. \eqref{e:4.1}, we applied the infinite product
form of the $\Theta_1$- and Dedekind
$\eta$-functions, and also their Jacobi
transformation formulas, e.g. see Ref. \cite{29}.

Now, by applying $\nu_0=i\upsilon_0$
in the amplitude \eqref{e:4.1}
we can obtain its imaginary part,
which exhibits the pair production of 
the open superstrings, and
hence the decay of the underlying system. Presence of the factor
$\sin{(\pi\upsilon_{0}t')}$ in the denominator
indicates the simple poles in 
the positive direction of the $t'$-axis,
i.e., $t'_k=k/\upsilon_0 $ where $k$ is any
positive integer number.
Hence, each pole separately leads to the creation
of an open superstring pair, and consequently
deterioration of the branes system. The decay rate
per unit volume of the D3-branes
is defined by $\mathcal{W}_{3,3}
=-2 {\rm Im}\mathcal{A}_{\rm annulus}/V_4$.
Thus, we obtain
\begin{eqnarray}
\mathcal{W}_{3,3}
&=&\frac{2 T_3^2 \pi^{d_{i_{\rm n}}/2}}
{\upsilon_0 V_4(2\pi)^{6}
\alpha'^{(d_{i_{\rm n}}-2)/2}}
\sqrt{\det ({\bf 1}-\mathcal{F}_1)
\det({\bf 1}-\mathcal{F}_2)}
\nonumber\\
&\times& \sum_k\bigg\{\left(-1\right)^{k+1}
\exp{\left(-k Y^2_{\rm n}/4\upsilon_0\alpha'\right)}
\left(\frac{k}{\upsilon_0}\right)^{(d_{i_{\rm n}}-8)/2}
\nonumber\\
&\times&\prod_{i_{\rm c}}
\Theta_3\left(\frac{y_{1}^{i_{\rm c}}
-y_{2}^{i_{\rm c}}}{2\pi R_{i_{\rm c}}}\bigg{|}
\frac{i\alpha'\upsilon_0}{k\pi R^2_{i_{\rm c}}}\right)
\bigg[ \sum_{\{N^{\alpha_{\rm c}}\}}
\left(2\pi\right)^{4}\prod_\alpha
\delta(p^\alpha_1-p^\alpha_2)
\nonumber\\
&\times&\exp{\left(-\frac{\upsilon_0}
{k\alpha'}l^{\alpha_{\rm c}}
l^{\beta_{\rm c}}\left(\delta_{\alpha_{\rm c}\beta_{\rm c}}+
\mathcal{F}^{\alpha}_{1 \alpha_{\rm c}}
\mathcal{F}_{2 \alpha\beta_{\rm c}}\right)
\right)}\bigg]
\frac{\sinh(\pi\upsilon_0)\sin(\pi\nu_1)}
{\sinh{(\frac{k\pi\nu_1}{\upsilon_0})}}
\nonumber\\
&\times&\left[\left(-1\right)^k
-\cosh{\left(\frac{k\pi\nu_1}{\upsilon_0}\right)}\right]^2
\mathcal{C'}_k(\upsilon_0,\nu_1)\bigg\},
\label{e:4.3}
\end{eqnarray}
in which the worldvolume of each D3-brane indicated by $V_4$,
and $\mathcal{C'}_k(\upsilon_0,\nu_1)$ has
the definition
\begin{eqnarray}
\mathcal{C'}_k(\upsilon_0,\nu_1)
&=&\prod^\infty_{m=1}
\frac{\left[1-2\left(-1\right)^k
\cosh(\frac{k\pi\nu_1}{\upsilon_0})
e^{-2m\pi k/\upsilon_0}+e^{-4m\pi k/\upsilon_0}\right]^4}
{\left(1-e^{-2m\pi k/\upsilon_0}\right)^6
\left[1-2\cosh(\frac{2k\pi\nu_1}{\upsilon_0})
e^{-2m\pi k/\upsilon_0}+e^{-4m\pi k/\upsilon_0}\right]}.
\label{e:4.4}
\end{eqnarray}

\subsection{An special case}

As we see, the above result is very complicated to accurately
express the influence of the parameters of
the setup on the decay rate. Thus, as an example,
we shall rewrite the foregoing rate for the
special matrices $\mathcal{F}_1$ and $\mathcal{F}_2$.
Therefore, let us choose the matrix
$\mathcal{F}_1$ $(\mathcal{F}_2)$, with the nonzero
electric field $E$ $(E')$ and magnetic field $B$ $(B')$
for the first (the second) D3-brane, as in the following
\begin{eqnarray}
\mathcal{F}^{\alpha}_{1\;\beta}=
\begin{pmatrix}
0 & -E & 0 & 0 \\
-E & 0 & 0 & 0 \\
0 & 0 & 0 & B \\
0 & 0 & -B & 0\\
\end{pmatrix} .
\label{e:4.5}
\end{eqnarray}
The eigenvalues $\lambda_0$
and $\lambda_1$ of the matrix $\Lambda$ satisfy
the equations
\begin{eqnarray}
\lambda_0+\lambda^{-1}_0
&=&\frac{2\left[1+4E E'+E'^2+E^2(1+E'^2)\right]}
{(1-E^2)(1-E'^2)},
\nonumber\\[10pt]
\lambda_1+\lambda^{-1}_1
&=&\frac{2\left[1+4B B'-2B^2-2B'^2(1-B^2)\right]}
{(1+B^2)(1+B'^2)} .
\label{e:4.6}
\end{eqnarray}
Thus, $\upsilon_0$ and $\nu_1$ possess the forms 
\begin{eqnarray}
\tanh(\pi\upsilon_0)&=&\frac{E+E'}{1+EE'},
\nonumber\\[10pt]
\tan(\pi\nu_1)&=&\frac{B'-B}{1+BB'}.
\label{e:4.7}
\end{eqnarray}

For this configuration, we receive
the decay rate as in the following
\begin{eqnarray}
\mathcal{W}_{3,3}
&=&\frac{2T_3^2 \pi^{d_{i_{\rm n}}/2}\;\sinh(\pi\upsilon_0)}
{(2\pi)^{6}\alpha'^{(d_{i_{\rm n}}-2)/2}}
\left(1+B^2\right)\left(1-E^2\right)
\sum_k\bigg\{\frac{1}{k}\left(-1\right)^{k+1}
\nonumber\\
&\times&\exp{\left(-kY^2_{\rm n}/4\upsilon_0\alpha'\right)}
\left(\frac{k}{\upsilon_0}\right)^{(d_{i_{\rm n}}-8)/2}
\prod_{i_{\rm c}}
\Theta_3\left(\frac{y_{1}^{i_{\rm c}}
-y_{2}^{i_{\rm c}}}{2\pi R_{i_{\rm c}}}\bigg{|}
\frac{i\alpha'\upsilon_0}{k\pi R^2_{i_{\rm c}}}\right)
\nonumber\\
&\times&\Omega\left(E,E',B,B',
\frac{\upsilon_0}{k},R_1,R_2,R_3\right)
\left[\left(-1\right)^k -1\right]^2
\mathcal{C'}_k(\upsilon_0,0)\bigg\},
\label{e:4.8}
\end{eqnarray}
\begin{eqnarray}
\Omega &=&\frac{\left(2\pi\right)^4}{V_4}
\sum^\infty_{n=-\infty}\sum^\infty_{s=-\infty}
\sum^\infty_{m=-\infty}\bigg\{
\delta[(E-E')n R_1/\alpha']
\nonumber\\
&\times&\delta[(B-B')s R_2/\alpha']
\delta[(B-B')m R_3/\alpha']
\nonumber\\
&\times&\exp{\left(-\frac{\upsilon_0}{k\alpha'}
\left[n^2 R^2_1(1-E^2)
+\left(s^2 R^2_2+m^2 R^2_3\right)(1+B^2)\right]
\right)}\bigg\}.
\label{e:4.9}
\end{eqnarray}
The $\Omega$-function comes from the third line of
Eq. \eqref{e:4.1}. According to the delta-functions
we applied $E=E'$ and $B=B'$. Hence, the second
equation of Eqs. \eqref{e:4.7} gives $\nu_1 =0$.
Note that $-1 <E<1$, and $\mathcal{C'}_k(\upsilon_0,0)$
is calculated from Eq. \eqref{e:4.4} in terms
of $\nu_1=0$. Besides, in this special
configuration, the tangential
directions $\{x^1, x^2, x^3\}$ are compact.

Now we compute the open superstring pair production rate.
According to \cite{30}, it is specified with the $k=1$ term.
Thus, we acquire
\begin{eqnarray}
\mathcal{W}^{(k=1)}_{3,3}
&=&\frac{8T_3^2 \pi^{d_{i_{\rm n}}/2}\sinh(\pi\upsilon_0)}
{(2\pi)^{6}\alpha'^{(d_{i_{\rm n}}-2)/2}}
\left(1+B^2\right)\left(1-E^2\right)
\exp{\left(-Y^2_{\rm n}/4\upsilon_0\alpha'\right)}
\nonumber\\
&\times&{\upsilon_0}^{(8-d_{i_{\rm n}})/2}\;\prod_{i_{\rm c}}
\Theta_3\left(\frac{y_{1}^{i_{\rm c}}
-y_{2}^{i_{\rm c}}}{2\pi R_{i_{\rm c}}}\bigg{|}
\frac{i\alpha'\upsilon_0}{\pi R^2_{i_{\rm c}}}\right)
\Theta_3\left(0\bigg{|}\frac{i \upsilon_0R^2_1
\left(1-E^2\right)}{\pi\alpha'}\right)
\nonumber\\
&\times&\Theta_3\left(0\bigg{|}\frac{i \upsilon_0R^2_2
\left(1+B^2\right)}{\pi\alpha'}
\right)\Theta_3\left(0\bigg{|}\frac{i \upsilon_0R^2_3
\left(1+B^2\right)}{\pi\alpha'}\right)
\mathcal{C'}_1\left(\upsilon_0  , 0\right).
\label{e:4.10}
\end{eqnarray}
For the very small electric fields
on the D3-branes, i.e. $|E|\ll 1$,
Eqs. \eqref{e:4.7} gives 
$\upsilon_0\approx 2|E|/\pi + \mathcal{O}(E^3)$.
This implies that $\upsilon_0\ll1$,
and hence $\mathcal{C'}_1(\upsilon_0 , 0)\approx1$.
Besides, for the finite values of the phrases  
$R^2_1$ and $R^2_{2,3}(1+B^2)$,
the four $\Theta_3$s in Eq. \eqref{e:4.10}
reduce to 1. Finally, in this limit,
the rate of the open superstring pair production
takes the form
\begin{eqnarray}
\mathcal{W}^{(k=1)}_{3,3}=
\frac{T_3^2 \pi^{d_{i_{\rm n}}/2}}
{8\pi^5 \alpha'^{(d_{i_{\rm n}}-2)/2}}\;
\left(1+B^2\right)
\left( \frac{2|E|}{\pi}\right)^{(10-d_{i_{\rm n}})/2}
\exp{\left(-\frac{\pi Y^2_{\rm n}}
{8\alpha' |E|}\right)}.
\label{e:4.11}
\end{eqnarray}
We observe that the pair creation rate has been prominently 
enhanced by the magnetic field. Though the two 
factors which contain the electric field are very small,
by increasing the magnetic field the rate can be
adjusted to any desirable value. In this case, the 
compactification radii $R_2$ and $R_3$ should be 
sufficiently small such that $|E|R^2_{2,3}(1+B^2) \ll 1$.
As we see, the presence of the 
electric fields drastically plays the main role in the
production of open superstring pairs, as expected.

Let us rewrite this equation in the non-compact
spacetime. Therefore, we should replace
$Y^2_{\rm n} \to Y^2 =\sum^9_{i=4}\left(
y^i_1 - y^i_2 \right)^2$ and $d_{i_{\rm n}}\to 6$.
Accordingly, we receive 
\begin{eqnarray}
\mathcal{W}^{(k=1)}_{3,3}\bigg{|}_{\rm non-compact}=
\frac{T_3^2}{8\pi^2 \alpha'^2}\;
\left(1+B^2\right)\left( \frac{2|E|}{\pi}\right)^2
\exp{\left(-\frac{\pi Y^2}{8\alpha' |E|}\right)}.
\label{e:4.12}
\end{eqnarray}
The exponential factors of Eqs. \eqref{e:4.11}
and \eqref{e:4.12} elaborate that by compactifying
some of the spatial directions the pair production rate
possesses an enhancement.

\section{Conclusions}

In the context of the superstring theory 
we introduced a boundary state, associated
with a D$p$-brane, in which some of the longitudinal and
transverse directions have been compacted on tori.
The brane has been dressed with electric
and magnetic fields. The closed superstring
amplitude was also presented. Afterward, we
changed the cylinder amplitude to the annulus one.

The 1-loop annulus amplitude enabled us
to compute the decay rate of two parallel D3-branes system
through the pair production of the open superstrings.
Since the background fields on the branes are
arbitrary and different, the foregoing rate
found a generalized form. However, for clarity
we chose special electric and magnetic fields on the branes.
Thus, we observed that for
small electric fields the pair production rate is
independent of the branes distance in the
compact subspace. In addition, in comparison with the
non-compact spacetime, this specific configuration
demonstrated that the compactification enhances the value
of the pair creation rate. Besides, the magnetic field 
also enhances the rate. Moreover,
for acquiring a nonzero pair production rate,
the electric fields are clearly
needed to induce a polarized region between the branes.



\begin{thebibliography}{99}

\bibitem{1}
J. Polchinski, Phys. Rev. Lett. \textbf{75}
(1995) 4724; J.~Polchinski, ``{\it String Theory}'',
(Cambridge University Press, Cambridge, 1998),
Volumes I and II. For a review, also see J. Polchinski,
hep-th/9611050.

\bibitem{2}
P. Di Vecchia, M. Frau, I. Pesando, S. Sciuto,
A. Lerda and R. Russo,  Nucl. Phys. \textbf{B 507}
(1997) 259-276.

\bibitem{3}
M. Billo, P. Di Vecchia and D. Cangemi,
Phys.Lett. \textbf{B 400} (1997) 63-70.

\bibitem{4}
M. Billo, P. Di Vecchia, M. Frau,
A. Lerda, I. Pesando, R. Russo and
S. Sciuto, Nucl.Phys. \textbf{B 526} (1998) 199.

\bibitem{5}
F. Hussain, R. Iengo and C. Nunez,
Nucl. Phys. \textbf{B 497} (1997) 205.

\bibitem{6}
O. Bergman, M. Gaberdiel and G. Liftschytz,
Nucl. Phys. \textbf{B 509} (1998) 194.

\bibitem{7}
P. Di Vecchia, A. Liccardo, in: YITP Workshop on
Developments in Superstring and
M Theory, Kyoto, Japan, 1999, in: YITP Proceedings Series,
vol. \textbf{4}, 1999;
P. Di Vecchia, M. Frau, A. Lerda,
A. Liccardo, Nucl. Phys. \textbf{B 565} (2000)397-426;
P. Di Vecchia, A. Liccardo, R. Marotta,
F. Pezzella, JHEP09 (2004) 050;
Int. J. Mod. Phys. \textbf{A 20} (2005) 5699-4795.

\bibitem{8}
C. Bachas, Phys. Lett. \textbf{B 374} (1996) 37.

\bibitem{9}
C.G. Callan, I.R. Klebanov, Nucl. Phys.
\textbf{B 465} (1996) 473.

\bibitem{10}
S. Gukov, I.R. Klebanov, A.M. Polyakov, Phys. Lett.
\textbf{B 423} (1998) 64.

\bibitem{11}
H. Arfaei and D. Kamani, Phys. Lett. \textbf{B 452}
(1999) 54-60, arXiv:hep-th/9909167;
D. Kamani, Phys. Lett. \textbf{B 487} (2000) 187-191, arXiv:hep-th/0010019;
E. Maghsoodi and D. Kamani,
Nucl. Phys. \textbf{B 922} (2017) 280-292, arXiv:1707.08383 [hep-th];
F. Safarzadeh-Maleki and D. Kamani,
Phys. Rev. \textbf{D 90} (2014) 107902, arXiv:1410.4948 [hep-th];
M. Saidy-Sarjoubi and D. Kamani,
Phys. Rev. \textbf{D 92} (2015) 046003, arXiv:1508.02084 [hep-th];
D. Kamani, Europhys. Lett. \textbf{57} (2002) 672-676, arXiv:hep-th/0112153.

\bibitem{12}
F.~Safarzadeh-Maleki and D.~Kamani,
Phys. Rev. \textbf{D 89} (2014) 026006, arXiv:1312.5489 [hep-th];
D. Kamani, Mod. Phys. Lett. \textbf{A 17} (2002) 237-243, arXiv:hep-th/0107184;
S. Teymourtashlou and D. Kamani,
Eur. Phys. J. C {\bf 81} (2021) 761, arXiv:2108.10164 [hep-th];
D. Kamani, Eur. Phys. J. C {\bf 26} (2002) 285-291, arXiv:hep-th/0008020.

\bibitem{13}
H. Arfaei and D. Kamani,
Nucl. Phys. \textbf{B 561} (1999) 57-76, arXiv:hep-th/9911146;
D. Kamani, Nucl. Phys. {\bf B 601} (2001) 149-168, arXiv:hep-th/0104089;
N. Barghi-Janyar and D. Kamani, Int. J. Mod. Phys. {\bf A 38} (2023) 
2350021, arXiv:2302.03659 [hep-th].

\bibitem{14}
D. Kamani, Ann. Phys. \textbf{354} (2015) 394-400, arXiv:1501.02453 [hep-th];
H. Arfaei and D. Kamani, Phys. Lett. \textbf{B 475} (2000) 39-45, 
arXiv:hep-th/9909079;
D. Kamani, Mod. Phys. Lett. \textbf{A 15} (2000) 1655-1664, 
arXiv:hep-th/9910043.

\bibitem{15}
A. Hanany and E. Witten, Nucl. Phys.
\textbf{B 492} (1997) 152.

\bibitem{16}
T. Sato, Nucl.Phys. \textbf{B 682} (2004) 117-149.

\bibitem{17}
T. Kitao, N. Ohta, J.G. ZhouPhys. Lett.
\textbf{B 428} (1998) 68-74.

\bibitem{18}
T. Takayanagi, JHEP \textbf{0002} (2000) 040.

\bibitem{19}
U.H. Danielsson, G. Ferretti, I.R. Klebanov,
Phys. Rev. Lett. \textbf{79} (1997) 1984-1987.

\bibitem{20}
U.H. Danielsson, G. Ferretti,
Nucl. Phys. Proc. Suppl. \textbf{68} (1998) 78-83.

\bibitem{21}
N. Ohta, T. Shimizu and J. Zhou,
Phys. Rev. \textbf{D 5} (1998) 2040-2044.

\bibitem{22}
T. Matsuo and T. Yokono, Mod. Phys. Lett.
\textbf{A  14} (1999) 1175-1182.

\bibitem{23}
Q. Jia, J. X. Lu, Z. Wu and X. Zhu, Nucl. Phys.
\textbf{B 953} (2020) 114947.

\bibitem{24}
C. Bachas and M. Porrati, Phys. Lett. \textbf{B 296} (1992) 77.

\bibitem{25}
C.P. Burgess, Nucl. Phys. \textbf{B 294} (1987) 427.

\bibitem{26}
M. Frau, I. Pesando, S. Sciuto, A. Lerda and R.
Russo, Phys. Lett.\textbf{B 400} (1997) 52.

\bibitem{27}
E. Bergshoeff, E. Sezgin, C.N. Pope and
P.K. Townsend, Phys. Lett. \textbf{B 188} (1987) 70.

\bibitem{28}
C.G. Callan, C. Lovelace, C.R. Nappi and S.A. Yost,
Nucl. Phys. \textbf{B 308} (1988) 221-284;
Nucl. Phys. \textbf{B 288} (1987) 525-550;
Nucl. Phys. \textbf{B 293} (1987).

\bibitem{29}
E.T. Whittaker and G.N. Watson,
``{\it A Course of Modern Analysis}'', Cambridge
University Press, doi: 10.1017/CBO9780511608759.

\bibitem{30}
A.I. Nikishov, Nucl. Phys. {\bf B 21} (1970) 346;
Sov. Phys. JETP {\bf 30}, 660 (1970).

\end{thebibliography}
\end{document}